%% file: kektc5.tex
\documentclass[fleqn,twoside]{article}
\usepackage{espcrc2}

\usepackage{graphicx}
\usepackage[figuresright]{rotating}

\newcommand{\AmS}{{\protect\the\textfont2
  A\kern-.1667em\lower.5ex\hbox{M}\kern-.125emS}}
\input{pre2.tex}

\title{
Physics potential of future long baseline neutrino
oscillation experiments with KEK-JAERI HIPA 
}

\author{Mayumi Aoki\address[MCSD]
{Theory Group, KEK, Tsukuba, Ibaraki 305-0801, Japan}%
        \thanks{This work was supported in part by the Grant-in-Aid for
Scientific Research from the Ministry of Education, Culture, Sports, Science
and Technology of Japan.
}
.}
       
\begin{document}

\begin{abstract}
We study physics potential of Very Long
Base-Line (VLBL) Neutrino-Oscillation Experiments with the
High Intensity Proton Accelerator, which will be completed by the
year 2007 in Tokai-village, Japan.
As a target, a 100 kton-level water-$\check {\rm C}$erenkov detector
is considered at 2,100 km away.
Assuming the pulsed narrow-band $\nu_\mu$ beams, we study
sensitivity of such experiments to the neutrino mass hierarchy, 
the mass-squared differences,
one CP phase and three angles 
of the lepton-flavor-mixing matrix.
We find that experiments at a distance 2,100 km can determine 
the neutrino mass hierarchy if the mixing matrix element
$|U_{e3}|$ is not too small.
The CP phase and $|U_{e3}|$ can be constrained
if the large-mixing-angle solution of the solar-neutrino deficit
is realized.

\vspace{1pc}
\end{abstract}

\maketitle

\section{Introduction}
In order to measure the neutrino oscillation parameters,
such as the neutrino mass-squared differences and 
the elements of the 3$\times$3 Maki-Nakagawa-Sakata (MNS) lepton flavor-mixing
matrix elements \cite{MNS}, 
various long base-line (LBL)
neutrino oscillation experiments are proposed.
In Japan, as a sequel to the K2K experiment, 
a new LBL neutrino oscillation experiment between the
High Intensity Proton Accelerator (HIPA) \cite{HIPA}
and the Super-Kamiokande (SK) 
with the base-line length of $L$=295 km 
has been proposed \cite{H2SK}.
The HIPA has a 50 GeV proton accelerator, 
which will be completed by the year 2007
in the site of Japan Atomic Energy Research Institute (JAERI)
as the joint project of KEK and JAERI.
It is expected to deliver
$10^{21}$ protons on target (POT) 
in one year.
The intensity of the neutrino beam in the $\sim$ 1 GeV range is
two orders of magnitude 
higher than that of the KEK beam for the K2K experiment.
The HIPA-to-SK experiment with 
$\langle E_\nu \rangle$= 1.3 GeV will measure
the larger mass-squared difference
with 3 $\%$ accuracy and
the mixing angle at about 1 $\%$ accuracy.

 In this report we 
examine physics potential of Very Long Base-Line
(VLBL) neutrino oscillation experiments with HIPA and
a huge neutrino detector \cite{BAND}
in Beijing, at about $L$=2,100 km away.
As a beam option,
the pulsed narrow-band $\nu^{}_{\mu}$ beam (NBB) is assumed.
For a target at $L$=2,100 km,
we consider a 100 kton water-$\check {\rm C}$erenkov detector
which is capable of 
measuring the $\nu^{}_{\mu}$ to $\nu^{}_{e}$ transition probability,
$P_{\nu_\mu \to \nu_e}$,
and the $\nu^{}_{\mu}$ survival probability,
$P_{\nu_\mu \to \nu_\mu}$.
We study
the sensitivity of such experiments to the neutrino mass hierarchy,
the mass-squared differences, the three angles and one CP phase of the
MNS matrix \cite{H2B}.

\section{Neutrino oscillation in the three-neutrino model }

The MNS matrix
has three mixing angles
and three phases in general.
We adopt the following parameterization
\bea
V
=
\bmaT
U_{e 1}    & U_{e 2}    & U_{e 3} \\
U_{\mu 1}  & U_{\mu 2}  & U_{\mu 3} \\
U_{\tau 1} & U_{\tau 2} & U_{\tau 3}  
\emaT
\bmaT
1 & 0 & 0 \\
0 & e^{i \varphi_2} & 0 \\
0 & 0 & e^{i \varphi_3} 
\emaT
.~~~~~
\nn
\eea
Two Majorana phases in the latter matrix, 
$\varphi_2^{}$ and $\varphi_3^{}$, 
do not contribute to the neutrino oscillation.
The former matrix 
can be parameterized by 
three mixing angles and one phase
just 
the same as the CKM matrix.
Because the present neutrino oscillation experiments
constrain directly the elements,
$U_{e2}$, $U_{e3}$, and $U_{\mu 3}$,
these three matrix elements are 
taken as
the independent parameters.
Without losing generality, we can take 
$U_{e2}$ and $U_{\mu 3}$ to be real and non-negative.
By allowing $U_{e3}$ to have the complex phase 
$
U_{e3}=|U_{e3}| e^{-i\delta_{_{\rm MNS}}} 
~~(0 \leq \delta_{_{\rm MNS}} < 2\pi )\,,
$
the four independent parameters are $U_{e2}, U_{\mu 3}, |U_{e3}|$
and $\delta_{_{\rm MNS}}$.
All the other matrix elements
are then determined by the unitary conditions.

\def\cU#1#2{U_{#1}^{#2}}
\def\cUm#1#2{\wt{U}_{#1}^{#2}}

The constraints on the MNS matrix elements and the mass-squared differences
are given by the atmospheric- and solar-neutrino and
the CHOOZ reactor experiments. 
An analysis of the atmospheric-neutrino data
from the SK experiment \cite{atm_SK} finds
$\sin^2 2\theta_{_{\rm ATM}}\sim (0.88-1.0)$ and
$\delta m^2_{_{\rm ATM}} ({\rm eV}^2) 
\sim (1.6-4.0) \times 10^{-3}.$
From the observations of the solar-neutrino deficit by
the SK collaboration \cite{solar_SK}, 
the MSW large-mixing-angle solution (LMA) is
preferred to
MSW small-mixing-angle solution (SMA),
MSW low-$\delta m^2$ solution (LOW) and
Vacuum Oscillation solution (VO).
The CHOOZ experiment \cite{CHOOZ} 
gives the constraint
$\sin^2 2\theta_{_{\rm CHOOZ}}< 0.1$ for
$\delta m^2_{_{\rm CHOOZ}} > \numt{1.0}{-3}\,{\rm eV}^2$.
From the above experiments,
the independent parameters in the MNS matrix
are obtained as \\
$\cU{\mu 3}{}=\sqrt{1-\sqrt{1-\sin^2 2\theta_{_{\rm ATM}}}
}\Big/\sqrt{2}\,, $ \\
$\cU{e2}{} = $\\[2mm]
$\sqrt{1-|\cU{e3}{}|^2
-\sqrt{\left(1-|\cU{e3}{}|^2\right)^2-\sin^22\theta_{_{\rm SOL}}}}
\Big/\sqrt{2}\,,$ \\
$\l|\cU{e3}{}\r| = \sqrt{
1-\sqrt{1-\sin^2 2\theta_{_{\rm CHOOZ}}}
}\Big/\sqrt{2} < 0.16 \,.
$\\
Here,
we have made the identification
$
\delta m^2_{_{\rm SOL}} =
 \left|\delta m^2_{12}\right| 
\ll
 \left|\delta m^2_{13}\right| = \delta m^2_{_{\rm ATM}}\,,
$
with $\delta m^2_{ij}=m^2_j-m^2_i$. 

Since all the above constraints are obtained from
the survival probabilities which are even-functions
of $\delta m^2_{ij}$,
there are four neutrino-mass hierarchy cases corresponding to the
sign of the $\delta m^2_{ij}$ as shown in \Fgref{cases}.
If the MSW effect is relevant for
the solar-neutrino oscillation, then the hierarchy cases
II and IV are 
not favored.
The hierarchy I (III) is called `normal' (`inverted') hierarchy,
which corresponds to $\delta m^2_{12} > 0$ and 
$\delta m^2_{13} > 0$ ( $\delta m^2_{13} < 0$).
From the Super-Nova 1987A observation,
the hierarchy I is favored against the hierarchy III \cite{SN1987A}.
\begin{figure}[htb]
\includegraphics[angle=-90,scale=0.35]{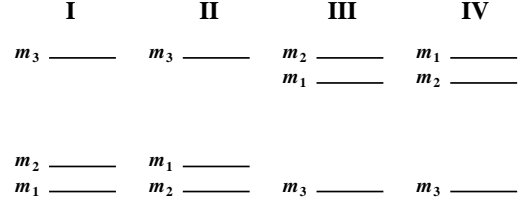}
\caption{Schematical view of the four cases of neutrino-mass hierarchy.}
\Fglab{cases}
\end{figure}


\section{VLBL neutrino oscillation experiment with $L$=2,100 km}

We explore the capability of a VLBL experiment with $L$=2,100 km
by using $\nu_\mu$ NBB.
The parameterization of the NBB is made 
to study the effects of changing the peak energy, $E_{\rm p}$,
of the NBB.
We find that the use of two different-energy NBB's, 
NBB's with $E_{\rm p}=$
4 GeV (NBB(4GeV)) and $E_{\rm p}=$ 6 GeV (NBB(6GeV)),
improves the physics
resolving power of the experiment significantly.
\Fgref{Pro_fl} shows
our parameterizations of 
the main part of NBB(4GeV) and NBB(6GeV) fluxes 
($\times$ neutrino-energy).
Also overlayed in \Fgref{Pro_fl} are 
the oscillation probabilities, which are calculated for 
$
\sin^22\theta_{_{\rm ATM}}=1.0,~
\delta m^2_{_{\rm ATM}}=3.5\times10^{-3} {\rm eV}^2, ~
\sin^22\theta_{_{\rm SOL}}=0.8,~
\delta m^2_{_{\rm SOL}}=10\times10^{-5} {\rm eV}^2, ~
\sin^22\theta_{_{\rm CHOOZ}}=0.1, ~
\delta_{_{\rm MNS}}=270^\circ
$
and the matter density,
$\rho =3$ g/cm$^3,$ with the hierarchy I.
Note that the NBB(6GeV) makes the $e$-like event
rates high
while keeping $\mu$-like event rates low. 

\begin{figure}[htb]
\vspace{9pt}
\includegraphics[scale=0.45]{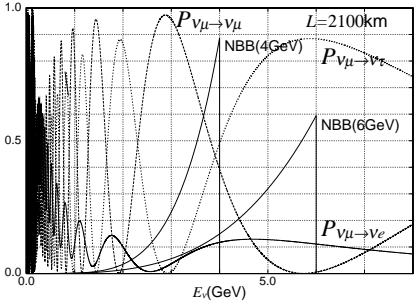}
\caption{
Overlayed are the NBB's (flux times $E_\nu$) used in our
analysis and
the neutrino oscillation probabilities at $L$=2,100 km.
}
\Fglab{Pro_fl}
\end{figure}


For a detector at $L$=2,100 km,
we assume
a 100 kton water-$\check {\rm C}$erenkov detector. 
It has a capability of distinguishing $e^\pm$ charged-current
(CC) events from $\mu^\pm$ CC events,
but does not 
distinguish their charges.
The detection efficiencies of the $\mu$-like and $e$-like
events are assumed to be 100 $\%$ for simplicity.
The statistical errors of each predictions are then simply square values
of the observed numbers of events.
We do not require capability of the detector to reconstruct the neutrino 
energy. 

The signals of our analysis are then the numbers of $\nu_\mu$ CC events
and those of $\nu_e$ CC events from the $\nu_\mu$ NBB.
For NBB($E_{\rm p}$), they are calculated as
\bea
&&N(\mu(e),E_{\rm p})= \nn \\
&&MN_A\int_0^{E_{\rm p}} dE_\nu \Phi(E_{\rm p}) 
\sigma^{\rm CC}_{\mu (e)}
P_{\nu_\mu \to \nu_{\mu (e)}} \,,~~~~ 
\nn
\eea 
where $M$ 
is the statistical significance
(mass of the detector (= 100 kton) times the running years),
$N_A=6.017\times 10^{23}$ is the Avogadro number,
$\Phi(E_\nu)$ is the flux 
at $L=2,100$ km.
The $\sigma^{\rm CC}_{\mu (e)}$ is the $\nu_\mu (\nu_e)$ CC cross
section per nucleon off water target.

\Fgref{main_B} shows 
the numbers of signal events, $N(\mu, E_{\rm p})$ and $N(e, E_{\rm p})$, 
expected at the base-line length of $L$=2,100 km for 500 kton$\cdot$year
with (a) the NBB(4GeV) and with (b) the NBB(6GeV).
The predictions for the four cases of the neutrino-mass hierarchies, 
I, II, III and IV, are indicated explicitly.
We take
\bea
&&
\sin^22\theta_{_{\rm ATM}}=1.0\,, ~
\delta m^2_{_{\rm ATM}}=3.5\times10^{-3} ~{\rm eV}^2\,~~~\nn
\\
&&\sin^22\theta_{_{\rm CHOOZ}}=0.06\,,0.1\,, \nn \\
&&\delta_{_{\rm MNS}}=0^\circ - 360^\circ\,,
\nn
\eea
with $\rho=3 ~{\rm g/cm^3}$.
The remaining two parameters,
$\sin^22\theta_{_{\rm SOL}}$ and $\delta m^2_{_{\rm SOL}}$, are taken
as the following sets for
the three possible solutions to the solar-neutrino deficit anomaly~:
$(\sin^22\theta_{_{\rm SOL}},\delta m^2_{_{\rm SOL}})
=(0.8, 15(3) \times 10^{-5}{\rm eV}^2)$ for LMA,
$(7\times 10^{-3}, 5 \times 10^{-6} {\rm eV}^2)$ for SMA,
$(0.7, 7\times10^{-11} ~{\rm eV}^2)$ for VO.

\begin{figure*}[htb]
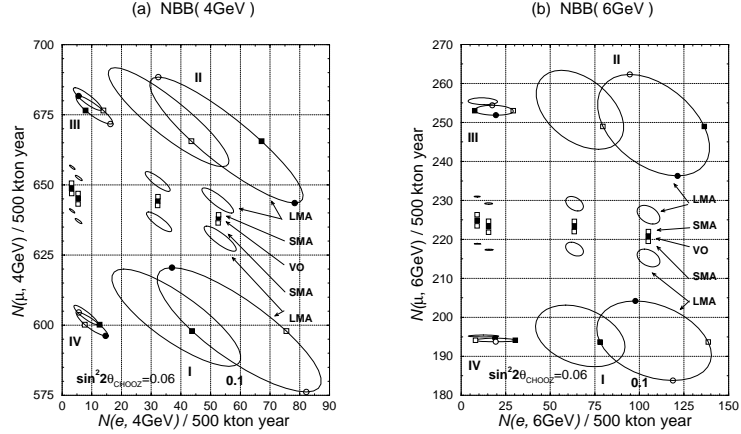

\begin{center}
\includegraphics[scale=0.25]{./figs/main4_1_06.eps}
\hspace{5ex}
\includegraphics[scale=0.25]{./figs/main6_1_06.eps}
\end{center}
\caption{%
The expected signal numbers, $N(\mu, E_{\rm p})$ and $N(e, E_{\rm p})$,
at $L$=2,100 km for 500 kton$\cdot$year with
(a) NBB(4GeV) and (b) NBB(6GeV).
The predictions for the four-cases of the neutrino-mass hierarchies
are depicted as I, II, III and IV.
The hierarchy cases II and IV are shown for completeness sake 
even though they do not lead to the MSW solution of the 
solar-neutrino deficit.  }
\Fglab{main_B}
\end{figure*}


For the hierarchy I,
the predictions 
reside in the corner at small $N(\mu, E_{\rm p})$ and large $N(e, E_{\rm p})$. 
For the LMA scenario, the predictions are shown by circles 
when $\delta_{_{\rm MNS}}$
is allowed to vary freely.
We show the four representative phase values by
solid-circle ($\delta_{_{\rm MNS}}=0^\circ$), solid-square ($90^\circ$),
open-circle ($180^\circ$), and open-square ($270^\circ$)
in the largest circle.
The two grand circles with smaller $N(\mu, E_{\rm p})$ give 
the predictions of the LMA solution
for $\delta m^2_{_{\rm SOL }}= 15\times 10^{-5}$ eV$^2$, and those with
larger
$N(\mu, E_{\rm p})$ are for $\delta m^2_{_{\rm SOL }}= 3\times 10^{-5}$
eV$^2$.
The $\nu_e$ CC event number
$N(e, E_{\rm p})$ grows with increasing $\sin^22\theta_{_{\rm CHOOZ}}$.
It is clearly seen that $\delta_{_{\rm MNS}}$ dependence 
is larger for larger $\delta m^2_{_{\rm SOL }}$ and for larger
$\sin^22\theta_{_{\rm CHOOZ}}$.
Note that the $\delta_{_{\rm MNS}}$ dependence of $N(e, 4{\rm GeV})$ 
is `orthogonal' to that
of $N(e, 6{\rm GeV})$.
The predictions of the SMA and the VO parameters appear 
just above the upper LMA circles,
where the $\delta_{_{\rm MNS}}$ dependence 
diminishes to zero.
We cannot distinguish the VO predictions between the
neutrino-mass
hierarchy I and II, nor between III and IV.

All the predictions of the hierarchy cases III and IV have very small 
$N(e, E_{\rm p})$, typically a factor of 5 or more smaller
in magnitude than the corresponding number of events in case I.
This striking sensitivity of $P_{\nu_\mu \to \nu_e}$
on the hierarchy cases is the bases of the capability of distinguishing
the cases in the VLBL experiments.
On the other hand, we find that the 5 $\%$ level differences in
$N(\mu, E_{\rm p})$ between the hierarchy cases I and III
are not useful for this purpose
because $P_{\nu_\mu \to \nu_\mu}$ depends strongly on 
the atmospheric parameters
$\delta m^2_{_{\rm ATM}}$ and
$\sin^22\theta_{_{\rm ATM}}$, and also
because $N(\mu, E_{\rm p})$ suffers from the neutrino-beam flux uncertainty.

We take into account for 
the background contributions to the $\mu$-like and the $e$-like events from
the secondary-beams
and the 
$\tau$ pure-leptonic decays.
The $e$-like events also receive contributions from the neutral-current
(NC) events where
produced
$\pi^0$'s mimic electron shower in the water-$\check {\rm C}$erenkov detector.
The error in the $e/{\rm NC}$ misidentification probability,
$
P_{e/{\rm NC}}=0.0055\pm 0.00055,
$
is accounted for as a systematic error.
The $\nu_\tau$ CC events with hadronic $\tau$-decays are counted as
NC-like events with the same $P_{e/{\rm NC}}$ factor. 
The 10 $\%$ errors in the branching fractions of the $\tau$ decay
are accounted for as systematic errors.

In addition, we account for 
the uncertainties in the total fluxes of the neutrino NBB's
and the matter density 
as the major
part of the systematic uncertainty.
We allocate common 3 $\%$ uncertainty for all the NBB's
and 3.3 $\%$ overall uncertainty in the matter density.

\section{Results of the $\chi^2$ analysis}
We examine the capability
of the VLBL experiments with $L=2,100$ km in determining the model parameters. 
The following questions are of our concern. \\ 
1. Can we distinguish the neutrino-mass hierarchy cases 
I ($\delta m^2_{13}>0$) and III ($\delta m^2_{13}<0$) ? \\ 
2. Can we measure $\sin^22\theta_{_{\rm CHOOZ}}$ and $\delta_{_{\rm MNS}}$ ? 
\\
3. Can we distinguish the solar-neutrino oscillation scenarios ? \\
4. How much can we improve the measurements of $\sin^22\theta_{_{\rm ATM}}$
and $\delta m^2_{_{\rm ATM}}$ ? \\ 

We examine the questions 
by combining two experiments with different base-line length,
$L=2,100$ km (HIPA-to-Beijing) and $L=295$ km (HIPA-to-SK).
For a VLBL experiment at $L=2,100$ km,
we assume the statistical significance of
500 kton$\cdot$year each with the NBB(4GeV)and NBB(6GeV).
As for the LBL experiment at $L$= 295 km, we assume that 100 kton$\cdot$year
for the low-energy NBB with $\langle p_\pi\rangle$= 2 GeV (NBB(2$\pi$)).
The data obtained from the LBL experiment is
what SK can gather in approximately 5 years with $10^{21}$ POT par year.
(The latest design of the HIPA-to-SK project \cite{KEKTC5} adopts 
off-axis beams in the first stage.)

The $\chi^2$ is a function of 
the three angles, the two mass-squared differences, the CP phase,
the flux normalization factors and the matter density.
For a given set of these parameters,
we have defined predictions of the $\mu$-like and $e$-like event numbers
including both the signal and the background.

\subsection{I v.s. III}
We show in \Fgref{IvsIII_B} the minimum $\chi^2$ as a function of 
$\sin^22\theta_{_{\rm CHOOZ}}$ by assuming the hierarchy III
in the analysis, when the expected event numbers
are generated for
the following input (true) parameters
with 
the hierarchy I in the LMA region. \\[3mm]
Input (true) parameters :\\
$ 
\begin{array}{ll}
\sin^22\theta_{_{\rm ATM}}^{true}=1.0,&
\delta m^{2~true}_{_{\rm ATM}}=3.5\times10^{-3} ~{\rm eV}^2\, 
\\
\sin^22\theta_{_{\rm SOL}}^{true}=0.8,&
\delta m^{2~true}_{_{\rm SOL}}=10\times10^{-5} ~{\rm eV}^2\, 
\end{array}
$
$
\sin^22\theta_{_{\rm CHOOZ}}^{true}=0.02\,, 0.04\,, 0.1\,, \\
\delta_{_{\rm MNS}}^{true}=0^\circ, 90^\circ, 180^\circ, 270^\circ\,.\\
\delta m^2_{12}=\delta m^{2~true}_{_{\rm SOL}},~~
\delta m^2_{13}=\delta m^{2~true}_{_{\rm ATM}} ({\rm hierarchy~I})$\\[3mm]
In total 12 minimum $\chi^2$ lines
are labeled by the input values of 
$\sin^22\theta^{true}_{_{\rm CHOOZ}}$ = 
0.02 (thin lines), 0.04 (medium-thick lines), 0.1 (thick lines) 
and 
$\delta^{true}_{_{\rm MNS}}$
= $0^\circ$ (solid lines), $90^\circ$ (long-dashed lines), 
$180^\circ$ (short-dashed lines), $270^\circ$ (dot-dashed lines). 

\begin{figure}[htb]
\vspace{9pt}
\includegraphics[scale=0.4]{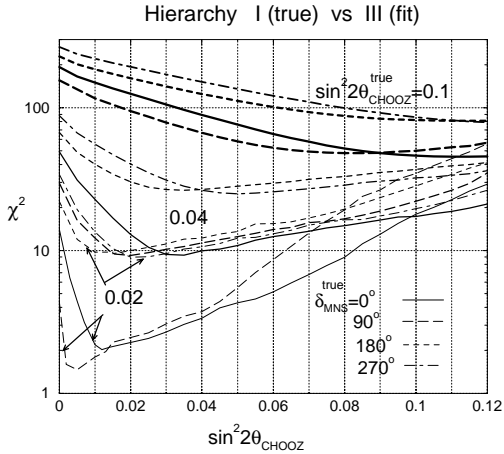}
\caption{Minimum $\chi^2$ as a function of  
$\sin^22\theta_{_{\rm CHOOZ}}$ when
the hierarchy III is postulated in the analysis.
The event numbers
are calculated for the hierarchy I in the LMA region. 
The input (true) values of $\sin^2\theta_{_{\rm CHOOZ}}$ 
and $\delta_{_{\rm MNS}}$ are 
shown by the 
line thickness and the line type, respectively.  }
\Fglab{IvsIII_B}
\end{figure}

The $\chi^2$ fit has been performed 
by assuming the hierarchy III.
The minimum of the $\chi^2$ function is found for a 
$\sin^22\theta_{_{\rm CHOOZ}}$ value in the range below 0.12, 
by varying 
the parameters, $\sin^22\theta_{_{\rm SOL}}$ and $\delta m^2_{_{\rm SOL}}$, 
in the LMA region, and
the remaining three parameters,
$\sin^22\theta_{_{\rm ATM}}$,
$\delta m^2_{_{\rm ATM}}$ and 
$\delta_{_{\rm MNS}}$, freely.\\[3mm]
Fitting parameters:\\
$
\begin{array}{ll}
\!\!\sin^22\theta_{_{\rm ATM}}~:~{\rm free},&
\!\!\delta m^2_{_{\rm ATM}}~:~ {\rm free}, \\
\!\!\sin^22\theta_{_{\rm SOL}}\!=0.7{\rm -}0.9,&
\!\!\delta m^2_{_{\rm SOL}}\!=(3{\rm -}15)\times10^{-5}{\rm eV}^2,~~~~  \\
\!\!\sin^22\theta_{_{\rm CHOOZ}}~:~{\rm free},&
\!\!\delta_{_{\rm MNS}}~:~ {\rm free}, 
\end{array}
$
$\delta m^2_{12}=\delta m^2_{_{\rm SOL}},
\delta m^2_{13}=-\delta m^2_{_{\rm ATM}} ({\rm hierarchy~III})$ \\[3mm]
The uncertainties in the total fluxes of the NBB's
and the matter density are taken into account.

The minimum $\chi^2$ for $\sin^22\theta^{true}_{_{\rm CHOOZ}}=0.1$
is always larger than 50, which allows us to
distinguish the hierarchy I from III at $7 \sigma$ level.
We also find that
the hierarchy III can be
rejected at 3$\sigma$ level 
when $\sin^22\theta_{_{\rm CHOOZ}} \gsim$ 0.04, and
at 1$\sigma$ level $\sin^22\theta_{_{\rm CHOOZ}} \gsim$ 0.02.
The capability of distinguishing the hierarchy cases I and III
comes from the remarkable differences of 
the transition probability $P_{\nu_\mu \to \nu_e}$ 
between them.

\subsection{$\sin^2 2\theta_{_{\rm CHOOZ}}$ and $\delta_{_{\rm MNS}}$}
 \Fgref{H2B_w_SK2} shows the minimum $\chi^2$
in the $\sin^2 2 \theta_{_{\rm CHOOZ}}$ v.s.
$\delta_{_{\rm MNS}}$ plane.
The event numbers are calculated at
$\sin^2 2\theta_{_{\rm CHOOZ}}^{true} = 0.06$ for 
$\delta_{_{\rm MNS}}^{true} = 0^\circ,90^\circ,180^\circ,270^\circ$
in each figure.
The other parameters are taken as the same as in \Fgref{IvsIII_B}. 
We show the input parameter point by a solid-circle
and the minimum $\chi^2=1, 4,$ and 9 by 
the solid-line, dashed-line and dotted-line, respectively.
The $\chi^2$ fit has been performed by assuming the LMA scenario
with the hierarchy I, while the other parameters are freely varied.

\begin{figure}[htb]
\vspace{9pt}
\includegraphics[scale=0.4]{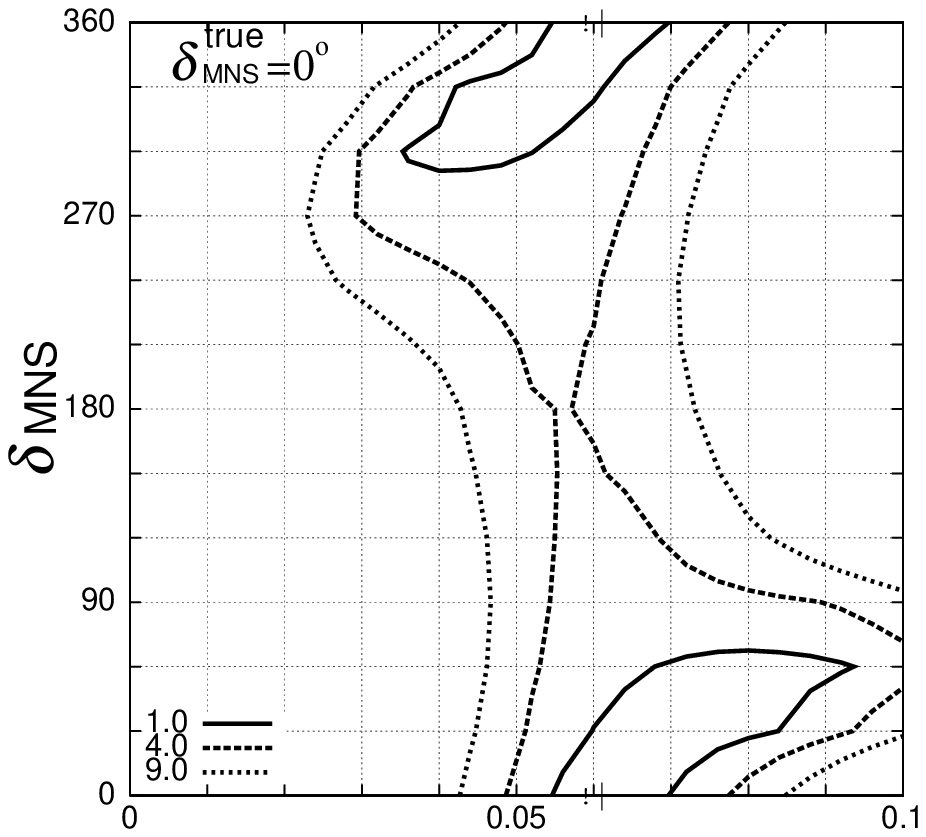}
\hspace{-5ex}
\includegraphics[scale=0.4]{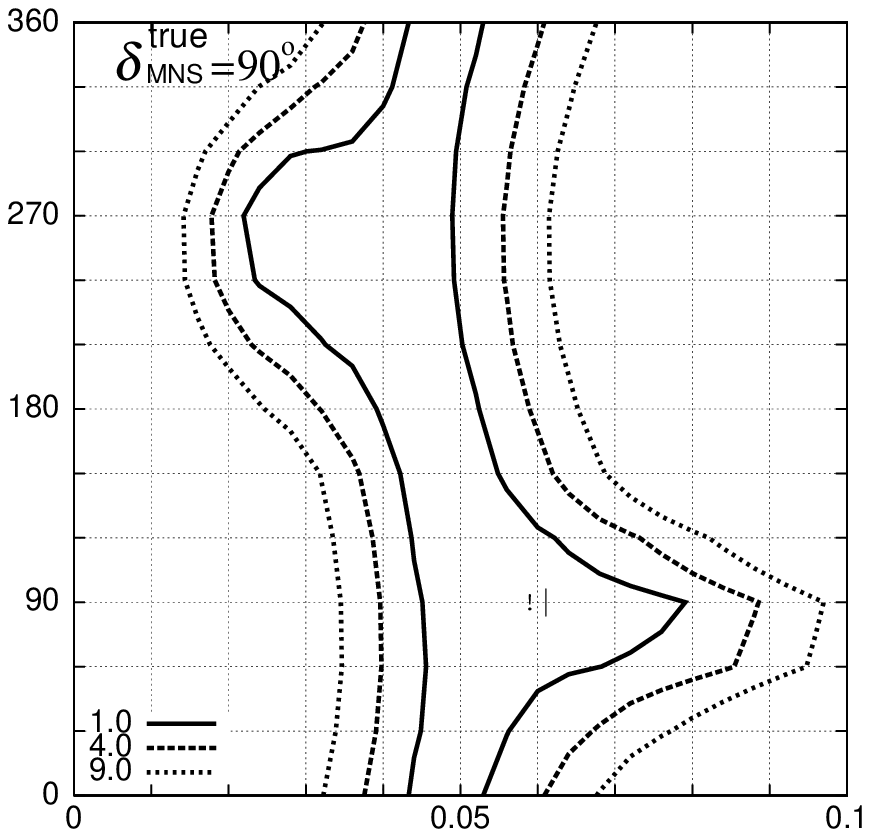}
\includegraphics[scale=0.4]{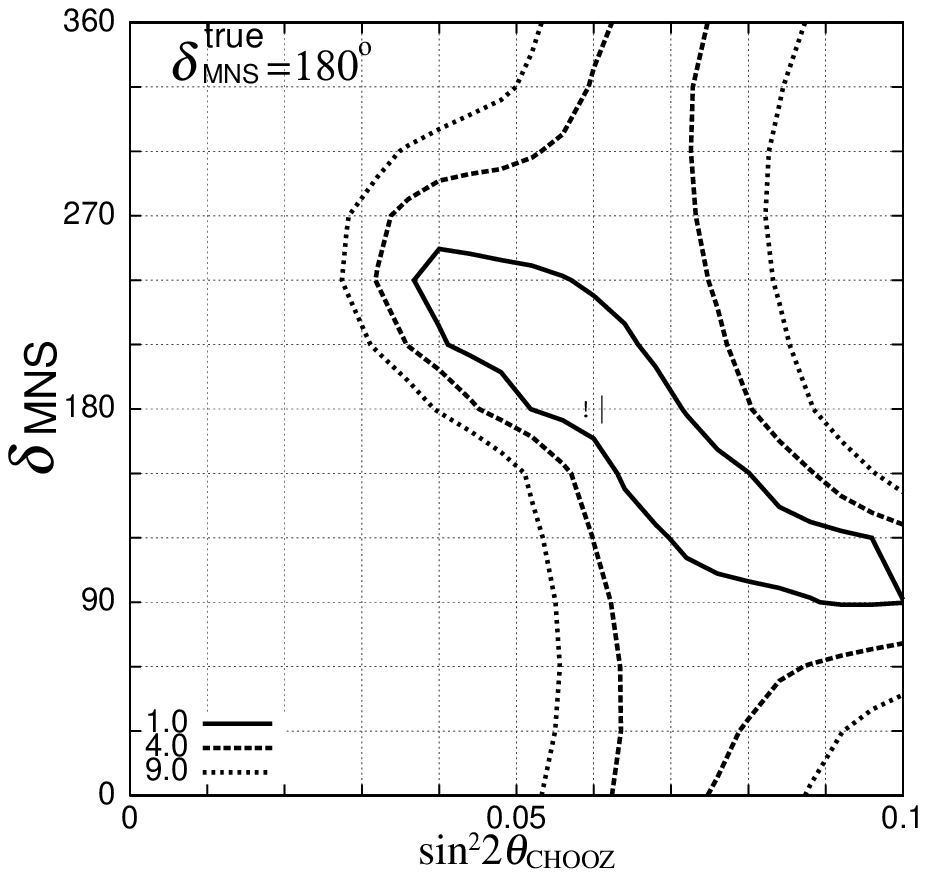}
\hspace{-2.5ex}
\includegraphics[scale=0.4]{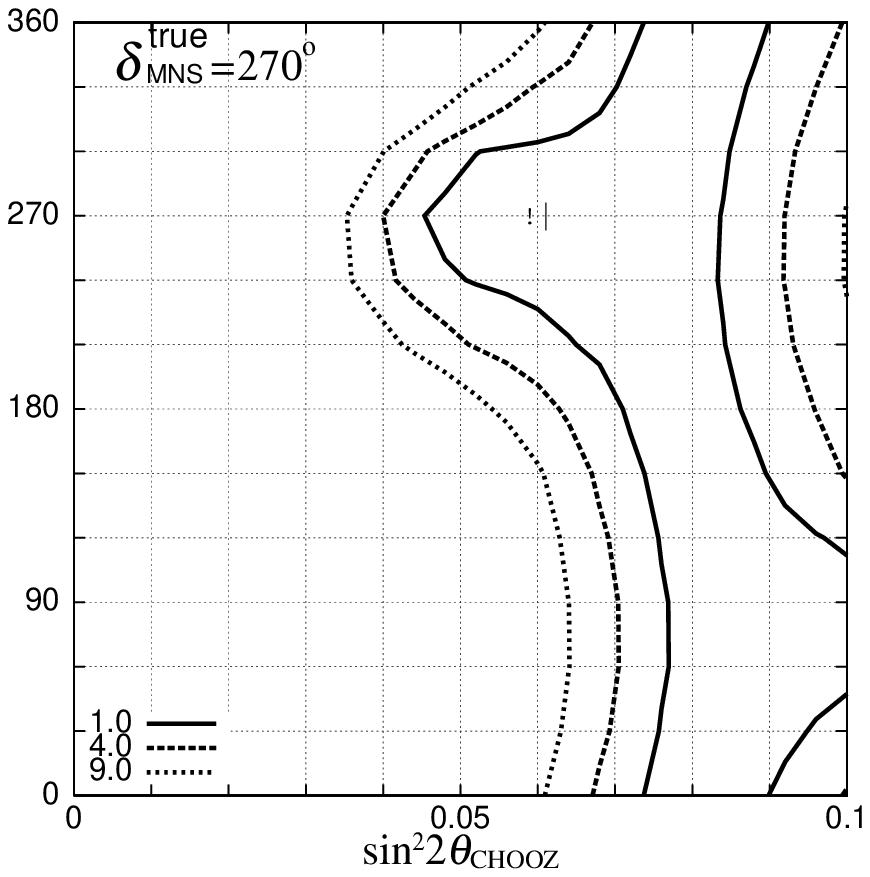}
\caption{
 Minimum $\chi^2$ contour plot
in the $\sin^2 2\theta_{_{\rm CHOOZ}}^{}$
v.s. $\delta_{_{\rm MNS}}^{}$ plane.
 The four values of the input parameters are given by the solid-circles.
 The $\chi^2$ fit has been performed by assuming the LMA scenario
with the hierarchy I.}
\Fglab{H2B_w_SK2}
\end{figure}

When $\delta^{true}_{_{\rm MNS}}$ is around $0^\circ$ or $180^\circ$,
$0.04 < \sin^22\theta_{_{\rm CHOOZ}} < 0.1$ is obtained and
$\delta_{_{\rm MNS}}$ can be constrained to local values at 1$\sigma$ level.
On the other hand, when
$\delta^{true}_{_{\rm MNS}}$ is around $90^\circ$ or $270^\circ$,
$0.02 < \sin^22\theta_{_{\rm CHOOZ}} < 0.08$  or
$0.045 < \sin^22\theta_{_{\rm CHOOZ}} < 0.12$ are obtained,
respectively, but we cannot constrain $\delta_{_{\rm MNS}}$.
As shown in \Fgref{main_B},
the $\delta_{_{\rm MNS}}=90^\circ$ ($270^\circ$) point 
for the hierarchy I 
lies in the lower (higher) $N(e, E_{\rm p})$ corner.
The same values of $N(e,4{\rm GeV})$ and $N(e,6{\rm GeV})$ 
can be obtained for 
the other $\delta_{_{\rm MNS}}$
by changing appropriately the $\sin^22\theta_{_{\rm CHOOZ}}$ parameter.
On the other hand, 
in the case of 
$\delta_{_{\rm MNS}}=0^\circ$ or $180^\circ$,
it is difficult to reproduce both
$N(e,4{\rm GeV})$ and $N(e,6{\rm GeV})$
simultaneously
because of the orthogonal
$\delta_{_{\rm MNS}}$-dependence between $N(e,4{\rm GeV})$ and
$N(e,6{\rm GeV})$.

\subsection{Solar- and atmospheric-neutrino
parameters
 }
We give only results for the remaining two questions.
If the LMA scenario is realized in Nature and 
$\sin^22\theta^{true}_{_{\rm CHOOZ}}\gsim 0.06$, 
the SMA/LOW/VO scenarios 
can be rejected at 1$\sigma$ level 
when $\delta^{true}_{_{\rm MNS}}$ is around
$0^\circ$ or $180^\circ$ but not at all
when $\delta^{true}_{_{\rm MNS}}$ is around
$90^\circ$ or $270^\circ$.

For the atmospheric-neutrino oscillation parameters,
$\sin^22\theta_{_{\rm ATM}}$ is measured to 1$\%$ level and
$\delta m^2_{_{\rm ATM}}$ with the 3$\%$ accuracy
when $\sin^22\theta^{true}_{_{\rm ATM}}=1.0$ and 
$\delta m^{2~true}_{_{\rm ATM}}=3.5 \times 10^{-3}$ 
with the LMA scenario.

%
%
\section{Summary}
The HIPA will be completed by the year 2007 and deliver $10^{21}$ POT
at 50 GeV for one year operation.
Assuming the NBB's from the HIPA and 
a 100 kton-level water-$\check {\rm C}$erenkov detector,
we study the physics potential of the VLBL experiments with $L$=2,100 km
as a sequel to the proposed HIPA-to-SK LBL experiment with $L=$295 km.
Thanks to the enhancement of matter effect, it is expected to 
distinguish the neutrino-mass hierarchy cases in such VLBL experiments.
We find that a combination of LBL experiments at $L$=295 km and VLBL
experiments at $L$=2,100 km can determine 
the neutrino mass hierarchy at 3$\sigma$ level if 
$\sin^22\theta_{_{\rm CHOOZ}} >$0.04.
Also it is found that, if the LMA scenario is realized in Nature,
$\sin^2 2 \theta_{_{\rm CHOOZ}}$ and $\delta_{_{\rm MNS}}$ can be 
constrained at 1$\sigma$ level in favorable cases.

\end{document}

%% file: pre2.tex
\def\beq{\begin{equation}}
\def\eeq{\end{equation}}
\def\bea{\begin{eqnarray}}
\def\eea{\end{eqnarray}}
\def\bseq{\begin{subequations}}
\def\eseq{\end{subequations}}
\def\nn{\nonumber}

\def\numt#1#2{#1 \times 10^{#2}}
\def\etal{{\it et al.}}

\def\btiny{\begin{tiny}}
\def\etiny{\end{tiny}}
\def\bsc{\begin{scriptsize}}
\def\esc{\end{scriptsize}}
\def\bfoot{\begin{footnotesize}}
\def\efoot{\end{footnotesize}}
\def\bsm{\begin{small}}
\def\esm{\end{small}}
\def\bno{\begin{normalsize}}
\def\eno{\end{normalsize}}
\def\bla{\begin{large}}
\def\ela{\end{large}}
\def\bLa{\begin{Large}}
\def\eLa{\end{Large}}
\def\bLA{\begin{LARGE}}
\def\eLA{\end{LARGE}}
\def\bhu{\begin{huge}}
\def\ehu{\end{huge}}
\def\bHu{\begin{Huge}}
\def\eHu{\end{Huge}}

\def\bCe{\begin{center}}
\def\eCe{\end{center}}
\def\bFR{\begin{flushright}}
\def\eFR{\end{flushright}}
\def\bFL{\begin{flushleft}}
\def\eFL{\end{flushleft}}

\def\PRL#1#2#3{Phys. Pev. Lett. {\bf #1}, #2 (#3)}
\def\PL#1#2#3{Phys. Lett. {\bf #1}, #2 (#3)}

\def\PTP#1#2#3{Prog. Theor. Phys. {\bf #1}, #2 (#3)}

\def\eqref#1{eq.(\ref{eqn:#1})}

\def\Fgref#1{Fig.~\ref{fig:#1}}

\def\Fglab#1{\label{fig:#1}}

\def\bmaT{\left(\begin{array}{ccc}}
\def\emaT{\end{array}\right)}
\def\bma{\left( \begin{array} }
\def\ema{\end{array} \right)}

\def\wt{\widetilde}
\def\l{\left}
\def\r{\right}
\def\gsim{~{\rlap{\lower 3.5pt\hbox{$\mathchar\sim$}}\raise 1pt\hbox{$>$}}\,}
\def\lsim{~{\rlap{\lower 3.5pt\hbox{$\mathchar\sim$}}\raise 1pt\hbox{$<$}}\,}